%% file: 0.main.tex
\PassOptionsToPackage{numbers,sort&compress}{natbib}

\documentclass[sigconf]{acmart}

\renewcommand\footnotetextcopyrightpermission[1]{} 
\usepackage{booktabs}
\usepackage{multirow}
\usepackage{amsmath}
\usepackage{enumitem}
\usepackage{tablefootnote}

\newcommand{\ie}{\textit{i.e.}}
\newcommand{\eg}{\textit{e.g.}}
\newcommand{\etc}{\textit{etc.}}

\newcommand{\datasettwo}{TencentGR-10M}
\newcommand{\challenge}{Tencent Advertising Algorithm Challenge 2025: All-Modality Generative Recommendation}

\acmConference {}{}{} 

\begin{document}


\title{The Tencent Advertising Algorithm Challenge 2025: \\ 
All-Modality Generative Recommendation}


\newcommand{\tencent}[1]{
  \affiliation{
    \institution{Tencent Inc.}
    \city{Shenzhen}
    \country{China}
  }
  \email{#1} 
}

\newcommand{\cuhk}[1]{
  \affiliation{
    \institution{CUHK}
    \city{Hong Kong SAR}
    \country{China}
  }
  \email{#1}
}

\author{
Junwei Pan\textsuperscript{\rm 1}, Wei Xue\textsuperscript{\rm 1}, Chao Zhou\textsuperscript{\rm 1}, Xing Zhou\textsuperscript{\rm 1}, Lunan Fan\textsuperscript{\rm 1}, Yanbo Wang\textsuperscript{\rm 1}, \\
Haoran Xin\textsuperscript{\rm 1}, Zhiyu Hu\textsuperscript{\rm 1}, Yaozheng Wang\textsuperscript{\rm 1}, Fengye Xu\textsuperscript{\rm 1}, Yurong Yang\textsuperscript{\rm 1}, Xiaotian Li\textsuperscript{\rm 1}, \\
Junbang Huo\textsuperscript{\rm 1}, Wentao Ning\textsuperscript{\rm 1}, Yuliang Sun\textsuperscript{\rm 2}, Chengguo Yin\textsuperscript{\rm 1}, Jun Zhang\textsuperscript{\rm 1}, Shudong Huang\textsuperscript{\rm 1}, \\
Lei Xiao\textsuperscript{\rm 1}, Huan Yu\textsuperscript{\rm 1}, Irwin King\textsuperscript{\rm 2}, Haijie Gu\textsuperscript{\rm 1}, Jie Jiang\textsuperscript{\rm 1}
}
\affiliation{%
    \textsuperscript{\rm 1}Tencent Inc. 
    \textsuperscript{\rm 2}CUHK\\
    \country{} 
  \{jonaspan,weixue,derekczhou\}@tencent.com
}

\renewcommand{\shortauthors}{Author et al.}

\begin{abstract}
Generative recommender systems are rapidly emerging as a new paradigm for recommendation, where collaborative identifiers and/or multi-modal content are mapped into discrete token spaces and user behavior is modelled with autoregressive sequence models. 
Despite progress on multi-modal recommendation datasets, there is still a lack of public benchmarks that jointly offer large-scale, realistic and fully all-modality data (including collaborative IDs, visual and textual modality features) designed specifically for generative recommendation (GR) in industrial advertising.
To foster research in this direction, we organised the \challenge, a global competition built on top of two all-modality datasets for GR: \textsc{TencentGR-1M} and \textsc{TencentGR-10M}. 
Both datasets are constructed from real de-identified Tencent Ads logs and contain rich collaborative IDs and multi-modal representations (text and vision) extracted with state-of-the-art embedding models. 
The preliminary track (\textsc{TencentGR-1M}) provides one million user sequences with up to 100 interacted items each, where each interaction is labeled with exposure and click signals, while the final track (\textsc{TencentGR-10M}) scales this to ten million users and explicitly distinguishes between click and conversion events at both the sequence and target level.

This paper presents the task definition, data construction process, feature schema, baseline generative recommendation model, evaluation protocol, and key findings from top-ranked and award-winning solutions. 
Our datasets focus on multi-modal sequence generation in an advertising setting and introduce weighted evaluation for high-value conversion events. 
We release our datasets at this link \footnote{https://huggingface.co/datasets/TAAC2025/TencentGR-1M \newline https://huggingface.co/datasets/TAAC2025/TencentGR-10M} and baseline implementations at this link \footnote{https://github.com/TencentAdvertisingAlgorithmCompetition/baseline\_2025} to enable future research on all-modality generative recommendation at an industrial scale.
The official website is https://algo.qq.com/2025.
\end{abstract}

\keywords{Recommender systems, Generative recommendation, Sequential recommendation, Multi-modal learning, Advertising, Competition, Dataset}

\maketitle

\input{1.intro}
\input{2.setting}

\input{3.dataset}

\input{4.baseline}

\input{5.platform}

\input{6.evaluation}

\input{7.outcome}

\input{8.conclusion}

\bibliographystyle{ACM-Reference-Format}
\bibliography{references}

\end{document}

%% file: 1.intro.tex
\section{Introduction}

Discriminative recommendation models have long been the dominant paradigm in industrial recommender systems. 
Their evolution has been marked by two major lines of progress: increasingly expressive feature interaction modeling~\cite{rendle2010fm,pan2018fwfm, sun2021fmfm,guo2017deepfm, huang2019fibinet, mao2023finalmlp, he2017nfm, lian2018xdeepfm, naumov2019dlrm, mahao2011isci, zhou2010userrec, cheng2016widedeep, zhang2022dhen, zhang2024wukong, guo2024collapse, pan2024ads_recommendation, zhu2025rankmixer} and increasingly powerful sequence-based user interest modeling~\cite{zhou2018din,zhou2019dien,feng2019dsin,zhou2024tin,pi2020sim,chang2023twin, feng2024long, chai2025longer, hu2025tencentlongseq, hou2024lcn, guo2025cain}. 
Building on these advances, recommender systems in large-scale platforms are now increasingly moving from discriminative formulations to generative architectures that operate directly on user behavior sequences~\cite{hou2025grsurvey,zhao2025grsurvey,wu2020mind,gao2022kuairand,yuan2022tenrec, zhou2025onerec, zhou2025onerecv2, xue2026generative, li2025leadremultifacetedknowledgeenhanced}. 

Instead of merely re-scoring a fixed candidate set, recent generative recommendation models~\cite{rajput2023tiger, badrinath2025pinrec, zhai2024HSTU, huang2025genrank} reformulate retrieval or ranking as sequence generation over item identifiers or semantic codes. 
These models focus on the following intertwined design axes: 
(i) how to organize the data which consists of both non-sequential tokens such as users' demographic features, as well as heterogeneous sequential tokens, including both the interacted item token and the action type token;
(ii) how to encode actions (\eg, exposure, click, conversion) as explicit tokens or conditioning signals so that different behavioral intents can be disentangled rather than collapsed into a single label.


In parallel, there is a rapidly growing body of work on integrating the extracted multi-modal representations into recommendation models. 
Beyond early semantic-ID-based models like TIGER~\cite{rajput2023tiger}, methods such as LETTER~\cite{wang2024letter}, DAS~\cite{ye2025das}, MMQ~\cite{xu2025mmq}, OneRec~\cite{zhou2025onerec} and parallel semantic IDs~\cite{zhang2024towards, hou2025rpg} design tokenisers that map multi-modal item content and collaborative signals into discrete code sequences suitable for generative retrieval and ranking. 
These works demonstrate that high-quality semantic IDs are crucial for bridging collaborative and content spaces and for scaling generative architectures to industrial corpora~\cite{zhang2025semanticids,spotify2025semanticids}.

Despite rapid methodological progress, the ecosystem of public benchmarks for generative recommendation is still limited. 
Most GR papers evaluate on medium-scale e-commerce corpora such as Amazon Beauty/Toys/Sports and Yelp, where items are represented by single-modality identifiers plus short textual metadata~\cite{rajput2023tiger,lee2025gram,paischer2024prefdisc,kong2025minionerec}.
Large-scale multi-modal recommendation datasets such as MIND for news~\cite{wu2020mind}, KuaiRand and KuaiRec for short
video~\cite{gao2022kuairand,gao2022kuairec}, Tenrec~\cite{yuan2022tenrec}, the WWW'25 short-video dataset with full video content~\cite{shang2025shortvideo}, and multi-modal user-interaction datasets from other domains~\cite{bruun2024mmuser} provide richer content, but they are typically designed for classic CTR or sequential recommendation and do not expose semantic IDs, industrial ad creatives, or conversion-centric labels tailored to GR.
Recent surveys~\cite{hou2025grsurvey,zhao2025grsurvey} on generative recommendation explicitly highlight the lack of large-scale, fully multi-modal, interactive benchmarks - especially in high-value industrial domains such as advertising - as a major bottleneck for the field.
In contrast, large-scale advertising platforms operate on long, heterogeneous and fully multi-modal user behavior sequences with both click and conversion signals, under strict privacy requirements.

To bridge this gap, we organised the Tencent Advertising 2025 All-Modality Generative Recommendation Challenge. 
The competition is centred on all-modality generative recommendation: given a user's all-modal ad interaction history, participants must predict the next ad they are most likely to interact with  (\eg, click or convert). 
Each interaction is represented by both collaborative identifiers (user and ad IDs, categories, \etc) and multi-modal embeddings distilled from the ad's text and visual creatives. 

%% file: 2.setting.tex
\section{Challenge Setting}

\subsection{Problem}
The core task of the challenge is a next-item recommendation problem on multi-modal ad interaction sequences. 
For each user $u$ we observe chronological sequence of ad-related behaviors (\eg, impressions, clicks, conversions):

\begin{equation}
\label{eq:seq}
    S_u = \{ x_u, x_{u,1}, x_{u,2}, \ldots, x_{u,T_u} \},
\end{equation}
where $x_u$ is a user-profile token aggregating static user features and each $x_{u,t}$ denotes an item token (an ad impression) at time $t$.

Particularly, each token is a tuple of rich side information:
\begin{equation}
\label{eq:token}
\begin{aligned} %
    &x_u = \big( f_{\mathrm{pf.}}^{(1)}, \ldots, f_{\mathrm{pf.}}^{(K_p)} \big), \\
    x_{u,t} = \big( &
    f_{\mathrm{cate}}^{(1)}, \ldots, f_{\mathrm{cate}}^{(K_a)};
    f_{\mathrm{act}};
    f_{\mathrm{mm}}^{(1)}, \ldots, f_{\mathrm{mm}}^{(K_m)} \big),
\end{aligned} 
\end{equation}
where $f_{\mathrm{pf.}}$ denotes user profile features (e.g., age, gender), $f_{\mathrm{cate}}$ denotes categorical attributes (e.g., ID, advertiser and product category), $f_{\mathrm{act}}$ denotes action/feedback signals (exposure, click, conversion), and $f_{\mathrm{mm}}$ denotes pre-computed embeddings produced by text and multi-modal encoders (item-only). Here, $K_p$, $K_a$, and $K_m$ denote the total number of user profile features, categorical attributes, and multi-modal embeddings, respectively.
Given a prefix of $S_u$, the goal is to generate the next ad $x_{u,T_u+1}$ that the user is most likely to interact with from a large candidate pool.

\subsection{Challenge Rounds}

The challenge is organized into two online rounds plus an on-site final round.

\paragraph{Preliminary round.}
Participants receive the \textsc{TencentGR-1M} dataset, containing approximately one million user sequences of impression and click behaviors.
The goal is to predict the next clicked ad.
They must submit training and inference code that runs inside our evaluation environment. 
A private test set is held out, and submissions are ranked by a weighted combination of HitRate@10 and NDCG@10. 
Only teams whose code can be successfully executed and reproduced are eligible for ranking. 
The top 50 teams advance to the second round.

\paragraph{Second round.}
Qualified teams receive the larger \textsc{TencentGR-10M} dataset with ten million user sequences and explicit click and conversion signals.
Compared with the preliminary round, \textsc{TencentGR-10M} logs conversion events inside the sequences and also treats impressions associated with conversions as valid prediction targets, making the task a \emph{next click-or-conversion prediction problem}. 
Participants again submit full code plus a technical report describing their method. 
Evaluation proceeds on a strictly black-box test set, with the similar top-$K$ metrics as in the preliminary round, but now incorporating behavior-type weighting to emphasize conversions, \ie, receiving higher scores when correctly predicting a converted ad(Section~\ref{sec:eval-second}). 
After code review and reproducibility checks, the top 20 teams are invited to the on-site final.

\paragraph{On-site final round}
Finalists present their methods and findings to the Challenge Committee. 
The overall ranking combines the final-round leaderboard score (75\%) and a committee review score (25\%) based on technical novelty, clarity, and potential impact.

\subsection{Awards and Talent Programs}
The challenge offers substantial incentives to encourage broad and high-quality participation. 
Specifically, the champion team receives a prize of 2{,}000{,}000 RMB, while the second and third place teams are awarded 600{,}000 RMB and 300{,}000 RMB, respectively. 
Teams ranked fourth to tenth each receive 100{,}000 RMB. 
In addition to the main rankings, we also establish a \emph{Technical Innovation Award} (200{,}000 RMB) to recognize outstanding originality in multimodal generative recommendation, further encouraging participants to explore new paradigms and breakthrough designs.

Beyond monetary prizes, all members of the finalist teams are eligible for a full-time offer interview opportunity with Tencent, and members of the top-10 teams are guaranteed to receive 
a formal full-time offer. 
Furthermore, all participants who advance to the second round are guaranteed an internship offer. 
For participants who are unable to onboard due to graduation timing or other academic constraints, Tencent will issue long-term offer intention letters on a case-by-case basis.

\subsection{Participation Rules and Target Audience}

The competition is open to full-time students worldwide, including undergraduate, master's, PhD and qualified postdoctoral researchers.
Each participant may join at most one team, and teams may consist of one to three members. 
Team formation and real-name verification are handled through the official competition website. 
Participants must ensure that their registration information is truthful and unique; use of multiple accounts or falsified identities leads to disqualification.

To emphasise methodological clarity and ensure fairness in comparison, we \emph{prohibit model ensembling} in our competition. 
This constraint is imposed not only to discourage heavy ensembling - which often shifts effort away from developing a sharp and well-designed single model - but also because large-scale ensemble systems are typically impractical in real-world industrial recommendation pipelines due to latency, memory and maintenance constraints.

Participants are also required to employ generative recommendation ideas, such as sequence modelling with autoregressive architectures or generative semantic ID construction, rather than purely relying on classic discriminative models. 

These rules are enforced through a manual inspection of the submitted training and inference code during the mandatory reproducibility check before promotion to the final round.

%% file: 3.dataset.tex
\section{Data}
\label{sec:data}


The \textsc{TencentGR} datasets are constructed from de-identified logs of Tencent Ads.
We first set an \emph{answer time window} $[t_\text{begin}, t_\text{end}]$.
Then we sample $N$ users who have a positive behavior, \ie, click in the preliminary round and click or conversion in the second round, in this window.
The ad corresponding to these positive behaviors in the answer time window is the prediction target or answer.
Each user's behaviors \emph{before the target exposure} are then treated as the sequence, which consists of:


\begin{itemize}[leftmargin=*]
  \item A \emph{user token} aggregating static and slowly changing profile features.
  \item A sequence of \emph{item tokens}, each corresponding to an ad exposure, click, or conversion, with rich side information including sparse IDs, categorical attributes, and multiple extracted text and multi-modal embeddings.
\end{itemize}

All personally identifiable information and raw creative content (\eg, images, videos, raw ad text) are removed.
Instead, we only expose hashed IDs and embedding vectors extracted with production models (Section~\ref{sec:multimodal}) to ensure privacy.
%
Table~\ref{tab:dataset-overview} summarises their basic statistics.
Details of the two datasets will be introduced in the following sections.

\begin{table}[t]
  \centering
  \caption{Overview statistics of the \textsc{TencentGR} datasets.}
  \label{tab:dataset-overview}
  \begin{tabular}{ccc}
    \toprule
    & \textsc{TencentGR-1M} & \textsc{TencentGR-10M} \\
    \midrule
    \# users & 1{,}001{,}845 & 10{,}139{,}575 \\
    \# ads & 4{,}783{,}154 & 17{,}487{,}676 \\
    max sequence length & 100 & 100 \\
    avg sequence length & 91.06 & 97.29 \\
    \# candidate ads & 660{,}000 & 3{,}637{,}720 \\
    \multirow{3}{*}{action type} & exposure (90.19\%) & exposure (94.63\%) \\
    & click (9.81\%) & click (2.85\%) \\
    & - & conversion (2.52\%) \\
    \bottomrule
  \end{tabular}
\end{table}

\subsection{Preliminary Round: \textsc{TencentGR-1M}}

\paragraph{Sequence and target construction.}
From Tencent Ads logs, we sample 1{,}001{,}845 users with at least one click in the answer time window. 
For each user, we locate the first clicked ad after $t_\text{begin}$ and attribute that click to its corresponding impression (exposure) event. 
The impression that triggered this first click is taken as the prediction target for that user. 
The observed history for the sequence consists of all user behaviors (ad exposures and any clicks) that happened strictly before the exposure time of this target impression.
We then prepend a user token and truncate the resulting sequence to at most 100 item tokens.



\paragraph{User and ad features.}
The user token aggregates several fields, such as demographics and long-term interests.
The item token contains creative-level identifiers, product metadata, and multiple multi-modal embedding features.
Table~\ref{tab:feature} summarizes the sparse feature schema for \textsc{TencentGR-1M}.

\paragraph{Action type.}
In \textsc{TencentGR-1M}, item tokens are labelled with $r_{u,t} \in \{0,1\}$ for exposure and click.
Clicks account for 9.81\% of all logged actions, while exposures account for 90.19\%.

\paragraph{Candidate set construction.}
During evaluation, each user is associated with a global candidate set of 660k de-duplicated ads.
We first ensure that every ground-truth target item is included in the candidate pool.
In addition, we sample non-target items from the ads corpus such that approximately 40\% of ads in the candidate pool appear as targets for at least one user. 
This balances the difficulty of retrieval and the diversity of negatives.

\begin{table}[t]
  \centering
  \caption{Feature schema and statistics in both \textsc{TencentGR-1M} and \textsc{TencentGR-10M}. "S" and "M" denote single-value and multi-value categorical features, respectively.}
  \label{tab:feature}
  \setlength{\tabcolsep}{1.2mm}{\begin{tabular}{cc|cc|cc}
    \toprule
    & & \multicolumn{2}{c|}{\textbf{\textsc{TencentGR-1M}}} & \multicolumn{2}{c}{\textbf{\textsc{TencentGR-1M}}} \\
    Feature ID & Type & \# values & Coverage & \# values & Coverage \\
    \midrule
    \multicolumn{6}{c}{\textbf{Ad Features}} \\
    100 & S & 6 & 99.93\% & 6 & 99.96\% \\
    101 & S & 51 & 99.93\% & 53 & 99.96\% \\
    102 & S & 90{,}709 & 99.03\% & 173{,}463 & 99.16\% \\
    112 & S & 30 & 97.94\% & 30 & 99.16\% \\
    114 & S & 20 & 99.91\% & 33 & 99.98\% \\
    115 & S & 691 & 32.02\% & 988 & 29.09\% \\
    116 & S & 18 & 99.90\% & 20 & 99.98\% \\
    117 & S & 497 & 97.94\% & 558 & 99.16\% \\
    118 & S & 1{,}426 & 97.92\% & 1{,}636 & 99.16\% \\
    119 & S & 4{,}191 & 97.90\% & 4{,}950 & 99.15\% \\
    120 & S & 3{,}392 & 97.80\% & 4{,}045 & 99.14\% \\
    121 & S & 2{,}135{,}891 & 100.00\% & 5{,}041{,}300 & 99.72\% \\
    122 & S & 90{,}919 & 99.93\% & 2{,}392 & 99.99\% \\
    
    \midrule
    \multicolumn{6}{c}{\textbf{User Features}} \\
    103 & S & 86 & 99.91\% & 87 & 99.96\% \\
    104 & S & 2  & 99.62\% & 2  & 99.76\% \\
    105 & S & 7  & 85.80\% & 7  & 85.67\% \\
    106 & M & 14 & 87.91\% & 14 & 88.25\% \\
    107 & M & 19 & 38.70\% & 19 & 35.89\% \\
    108 & M & 4  & 17.04\% & 4  & 15.36\% \\
    109 & S & 3  & 99.96\% & 3  & 99.98\% \\
    110 & M & 2  & 42.98\% & 2  & 40.51\% \\
    
    \bottomrule
  \end{tabular}}
  \end{table}

\subsection{Second Round: \datasettwo{}}

\paragraph{Sequence and target construction.}
For the final round we scale up the dataset by an order of magnitude and \emph{incorporate conversions}. 
We sample 10{,}139{,}575 users with at least one click or conversion. 
For each user, we again draw a reference time $t_\text{begin}$ and then search for the earliest qualifying target event after $t_\text{begin}$. 
If we can find a conversion, we first associate the conversion with the click that triggered it and then propagate the attribution to the underlying impression.
Otherwise, if we can only find a click, we attribute it to its triggering impression exactly as in the preliminary round. Note that we have set an attribution window consistent with industry standards to account for conversion delay.
In both cases, the impression associated with the first post-$t_\text{begin}$ conversion or click is treated as the prediction target. 
The observed history for that user consists of all logged behaviors (including exposures, clicks and conversions) that occurred strictly before the exposure time of the target impression, and the sequence is truncated to at most 100 item tokens.

Notably, \emph{conversions appear both as events within the user sequence and as part of the prediction target type}, which is a key difference compared with the preliminary round.

\paragraph{User and ad features.}
The feature schema largely follows \textsc{TencentGR-1M}, but with expanded cardinalities due to the much larger scale.
Table~\ref{tab:feature} reports the key statistics.

\paragraph{Action type.}
The second-round dataset adds conversion labels. All item tokens are labelled with $r_{u,t} \in \{0,1,2\}$ for exposure, click and conversion.
Among all logged actions we observe 94.63\% exposures, 2.85\% clicks, and 2.52\% conversions.
Although conversions are rare, they are significantly more valuable than clicks; we account for this in the evaluation by assigning higher relevance weights to conversions.

\paragraph{Candidate set construction.}
The global candidate set for the second round contains 3{,}637{,}720 ads.
We follow the same construction as in the preliminary round, ensuring that ground-truth targets are all included in this set.
We also include some randomly sampled ads from the whole data log to simulate the real-world scenario.

\subsection{Multi-modal Features}
\label{sec:multimodal}

\begin{figure}
    \centering
    \includegraphics[width=0.9\columnwidth]{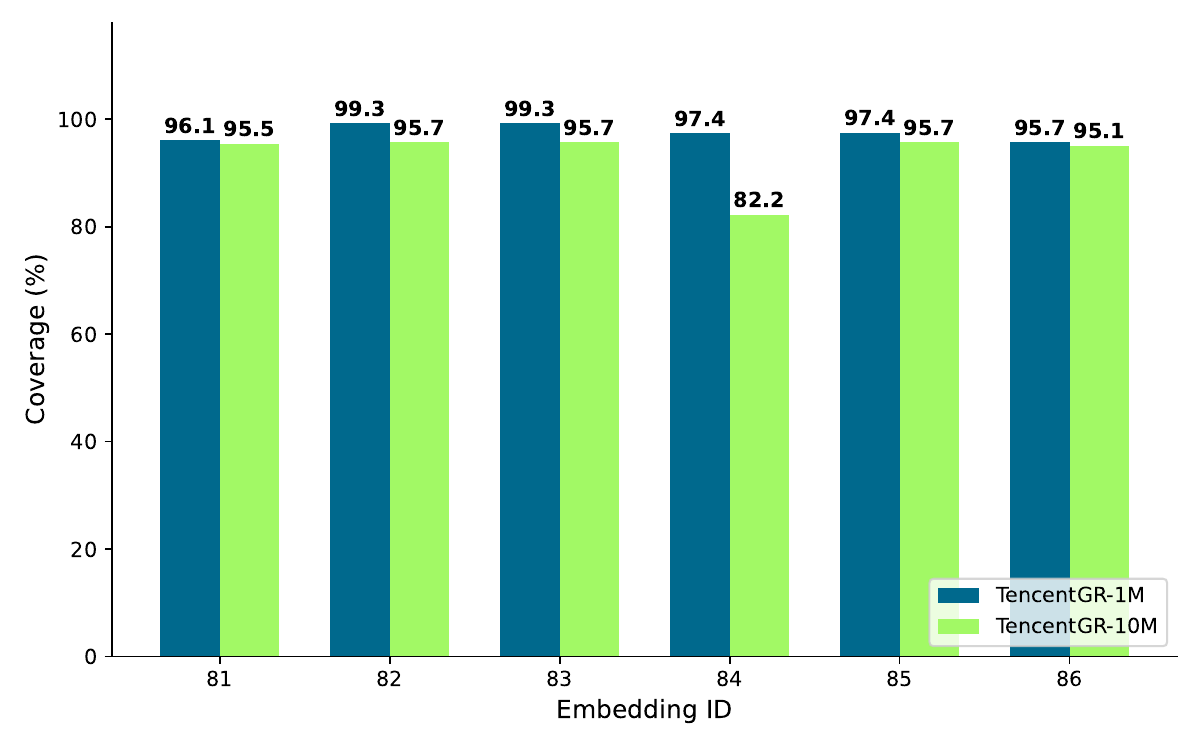}
    \caption{Coverage of the multi-modal embeddings on two datasets.}
    \label{fig:mm_coverage}
\end{figure}

Raw ad creatives include text (titles, descriptions), images and sometimes videos.
To protect advertiser privacy and reduce storage and bandwidth, we do not release raw creatives.
Instead, we extract multi-modal embeddings using a suite of production models.
Table~\ref{tab:mm-models} lists the models and their output dimensions.
Each creative may have up to six embedding vectors attached, corresponding to different encoders and modalities.
The Bert- and Hunyuan-finetune denotes that we finetune the Bert/Hunyuan model by our real-world collaborative data with a contrastive learning loss, and then extract the multi-modal embedding from the finetuned model, while all other models are not finetuned at all.
We further summarise their coverage in Figure~\ref{fig:mm_coverage}.

\begin{table}[t]
  \centering
  \caption{Multi-modal embedding models used to construct TencentGR. "T" denotes text and "I" denotes image.}
  \label{tab:mm-models}
  \setlength{\tabcolsep}{0.4mm}{\begin{tabular}{ccccc}
    \toprule
    Emb. ID & Model name & Modality & Params & Dim. \\
    \midrule
    81 & Bert-finetune              & T       & 0.3B & 32 \\
    82 & Conan-embedding-v1\cite{li2024conanembeddinggeneraltextembedding}      & T       & 0.3B & 1{,}024 \\
    83 & gte-Qwen2-7B-instruct~\cite{li2023gte-qwen}    & T       & 7B & 3{,}584 \\
    84 & hunyuan\_mm\_7B\_finetune  & I & 7B   & 4,096/32\tablefootnote{We use 4,096-dimensional embeddings for \textsc{TencentGR-1M} and 32-dimensional embeddings for \textsc{TencentGR-10M}.} \\
    85 & QQMM-embed-v1~\cite{xue2025qqmm}            & I & 8B   & 3{,}584 \\
    86 & UniME-LLaVA-OneVision-7B~\cite{gu2025unime} & I & 8B   & 3{,}584 \\
    
    \bottomrule
  \end{tabular}}
\end{table}

%% file: 4.baseline.tex
\section{Baseline Model}

To provide a strong reference implementation and lower the barrier to entry, we release a baseline generative recommendation pipeline that participants can directly build upon. 
The baseline adopts a next-token prediction formulation with a causal Transformer backbone and approximate-nearest-neighbour (ANN) based retrieval.

\subsection{Training}

\paragraph{Sequence construction.}
For each user, we construct the input sequence as introduced in Equation~\ref{eq:seq}, where the first token aggregates all user-level features and each subsequent token represents one ad interaction. 
Each token consists of multiple feature fields drawn from the shared feature schema, together with multi-modal embeddings.

\paragraph{Feature encoding.}
We adopt a multi-field feature fusion design based on the token schema in Equation~\ref{eq:token}.
Each categorical/ID feature field corresponds to its own embedding table, while multi-modal features directly use the provided continuous embeddings.
For the user-profile token and each item-interaction token, we first perform field-wise embedding lookup, concatenate all field embeddings, and then apply a small feed-forward network to project them into the final token embedding space:
\begin{equation}
\label{eq:feature-encoding}
\begin{aligned}
    \mathbf{e}_f &= \mathrm{Emb}_f\!\left(f\right), \quad \forall f \in \mathcal{F}, \\
    \mathbf{x}_u^{0} &= \mathrm{MLP}\!\left(\operatorname{concat}\!\left(\{\mathbf{e}_f\}_{f \in \mathcal{F}_u}\right)\right), \\
    \mathbf{x}_{u,t}^{0} &= \mathrm{MLP}\!\left(\operatorname{concat}\!\left(\{\mathbf{e}_f\}_{f \in \mathcal{F}_{u,t}}, \{f_{\mathrm{mm}}^{(j)}\}_j\right)\right),
\end{aligned}
\end{equation}
where $\mathrm{Emb}_f(\cdot)$ denotes a learnable embedding table for the corresponding sparse field $f$, $\mathcal{F}_u$ is the set of user profile features, $\mathcal{F}_{u,t}$ is the set of item features (i.e., collaborative IDs and categorical attributes), and $f_{\mathrm{mm}}^{(j)}$ is the pre-computed continuous multi-modal embedding (thus used directly).
The resulting $\mathbf{x}_u^{0}$ and $\mathbf{x}_{u,t}^{0}$ are the final token representations fed into the Transformer backbone.

\paragraph{Backbone architecture.}
The sequence encoder is a causal Transformer. 
Given the token embeddings from Equation~\ref{eq:feature-encoding}, we first prepend the user-profile token and add positional encodings to preserve temporal order:
\begin{equation}
\label{eq:input-seq}
    \mathbf{H}^{0} = \left[ \mathbf{x}_u^{0} + \mathbf{p}_0, \; \mathbf{x}_{u,1}^{0} + \mathbf{p}_1, \; \ldots, \; \mathbf{x}_{u,T_u}^{0} + \mathbf{p}_{T_u} \right],
\end{equation}
where $\mathbf{p}_t$ denotes the learnable positional embedding at position $t$.
We then apply $L$ stacked Transformer layers with causal masking, so that the representation at position $t$ only attends to positions $\leq t$:
\begin{equation}
\label{eq:transformer}
    \mathbf{H}^{l} = \mathrm{TransformerLayer}^{l}\!\left(\mathbf{H}^{l-1}\right), \quad l = 1, \ldots, L.
\end{equation}
The final user state at position $t$ is taken as the last-layer hidden state:
\begin{equation}
\label{eq:user-state}
    \mathbf{h}_{u,t} = \mathbf{H}^{L}[t],
\end{equation}
which serves as the user embedding for predicting the next item at position $t+1$.

\paragraph{Training objective}
We formulate the problem as next-token prediction task. 
For each training instance $(u,t)$, we take the user state after processing the history up to position $t$ and treat the impression immediately after $t$ as the positive next item $i^+$. 
We then sample a set of negative items $\mathcal{N}_{u,t}$ from the global item pool uniformly among the whole candidate pool. 
The model outputs a score $s_{u,t, i}$ between the user state and each candidate item $i$.
We train the model using an InfoNCE loss~\cite{oord2018representation}:
\[
\mathcal{L}
= - \sum_{(u,t)}
\log
\frac{\exp(s_{u,t,i^+})}
{\exp(s_{u,t,i^+}) + \sum_{i^- \in \mathcal{N}_{u,t}} \exp(s_{u,t,i^-})}.
\]
This objective encourages the positive target to be scored higher than a randomly sampled set of negatives, and is naturally aligned with the retrieval-style evaluation used in the challenge. 
In the second-round baseline, we apply action-type weights to the loss to emphasize conversion events.

\[
\mathcal{L}
= - \sum_{(u,t,a)} w_a \cdot 
\log
\frac{\exp(s_{u,t,i^+})}
{\exp(s_{u,t,i^+}) + \sum_{i^- \in \mathcal{N}_{u,t}} \exp(s_{u,t,i^-})}.
\]

where $w_a$ denotes the loss weight for action type $a$.

\paragraph{Implementation Details}
We implement the baseline model with $1$ transformer block and set the hidden dimension to $d=32$. The number of attention heads is set to $1$, and the dropout rate is fixed at $0.2$. For each user, the historical behavior sequence is truncated or padded to a maximum length of $101$. Item embeddings and positional embeddings are jointly learned. We train the model using the Adam optimizer with learning rate $0.001$. Following the standard training protocol for sequential recommendation, for each positive target item we sample one negative item from the whole item vocabulary. The model is implemented in PyTorch and trained on a single high-performance GPU. Additional implementation details and hyperparameter settings are provided in the released code.

\subsection{Inference}

At inference time, user representation learning and item retrieval are decoupled:

\begin{itemize}
    \item \textbf{User embedding.} We feed a user's behavior history into the Transformer and take the last-layer hidden state at the final position as the user embedding, which summarizes recent behavior and context.

    \item \textbf{Candidate Item embedding.} For each candidate item in the candidate pool, we apply the same feature encoder used during training to obtain an item embedding. These embeddings can be pre-computed and cached.

    \item \textbf{Approximate nearest neighbor search.} We build an ANN index over all item embeddings. 
    At serving time, the user embedding is used as a query to retrieve the top-$K$ nearest items using Faiss~\cite{douze2024faiss}.
\end{itemize}

%% file: 5.platform.tex
\section{Competition Platform}

\begin{figure*}
    \centering
    \includegraphics[width=0.8\linewidth]{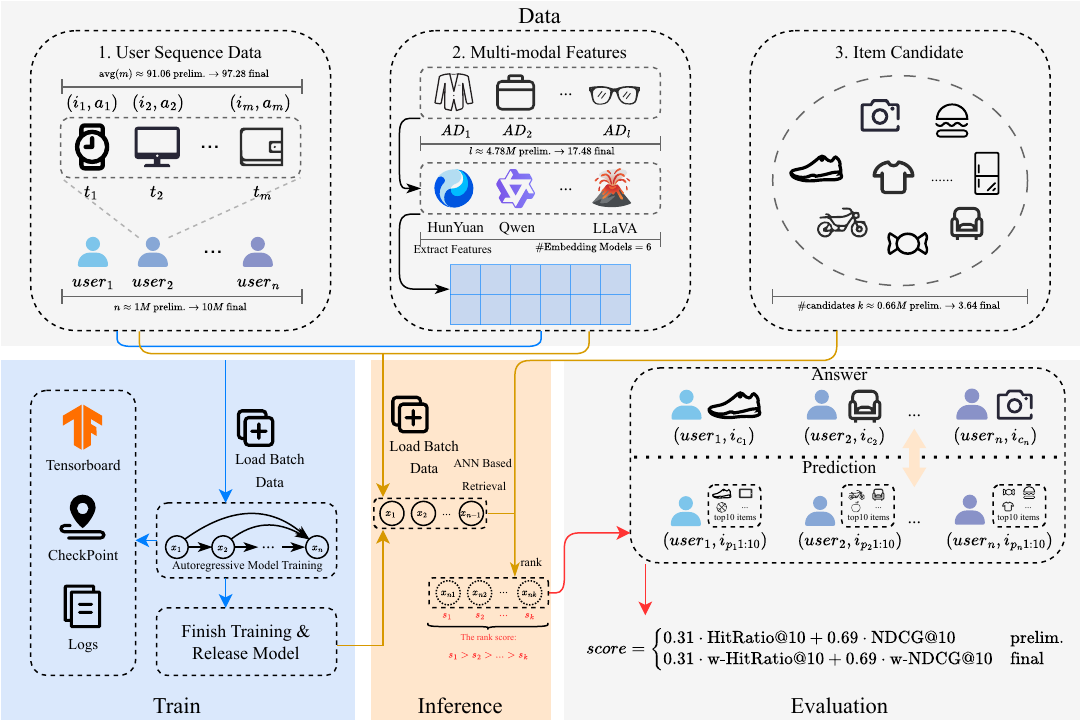}
    \caption{Illustration of the whole framework of the competition. The \textit{prelim.} denotes the preliminary round and the \textit{final} denotes the final round. The $\#$ denotes the number. The Data is divided into three parts, the first part is User Sequence Data where $(i_j, a_j)$ denotes the item and action at time $t_j$, $1<=j <=m$; the second part is Multi-modal Features, as explained in \ref{sec:multimodal}, the multi-modal features are extracted from 6 embedding models; the third part is the item candidates. The process of the competition is also divided into three parts: Autoregressive training, inference and evaluation, where the first two parts are completed by participants. After inference, the model will produce the top-k prediction items, then we will compute the final score according to the prediction and ground-truth answer in the evaluation stage. }
    \label{fig:main_fig}     
\end{figure*}


The overall process and date for the challenge are shown as Figure~\ref{fig:main_fig}. 
All competition workflows are executed on the Tencent Angel machine learning platform~\cite{zhao2024efficiently, nie2023angel}. 
Angel provides distributed training and evaluation capabilities designed for large-scale advertising models. 
For this challenge, we exposed the following functionalities to participants:

\begin{itemize}

    \item \textbf{Reference implementations:} we provide scripts for data loading, baseline training and evaluation on both \textsc{TencentGR-1M} and \textsc{TencentGR-10M}.

    \item \textbf{Virtualised GPU resources:} Angel offers fine-grained resource allocation via virtual GPU cards.  
    Participants can submit jobs using as fine-grained as 0.2 of a physical GPU, enabling efficient hardware sharing.
    We provide up to 0.2 and 7 high-performance GPUs for each team in the preliminary and second round during the competition.
    \item \textbf{High-throughput execution:} throughout the competition, Angel executed hundreds of thousands of training and evaluation jobs while maintaining stable service.
\end{itemize}

The evaluation environment for test data is strictly sandboxed and fully isolated from user training environments. 
Participants submit inference code that reads from the provided test set and produces predictions in a predefined JSON format. 
The evaluation container has no network access and cannot write outside a designated output directory, preventing leakage of test labels or execution of unauthorised programs.

Teams are allowed up to three submissions per 24-hour window. 
Each day, the public leaderboard is refreshed using each team's best score from the previous 24 hours. 
Repeated or severe violations - such as exploiting the platform, using unauthorised data or bypassing submission limits - may result in score invalidation or disqualification.

%% file: 6.evaluation.tex
\section{Evaluation Protocol}
\label{sec:eval}

\subsection{Preliminary-Round Evaluation Metrics}

In the preliminary round, the task is formulated as next-click prediction, where only click events are regarded as relevant signals.
To maintain simplicity and comparability with classic sequential recommendation benchmarks, we adopt the standard HitRate@$K$ and NDCG@$K$ without any behavior-type weighting.

Let $G_u$ denote the ground-truth next clicked item for user $u$, and let $\hat{y}_{u,1}, \ldots, \hat{y}_{u,K}$ be the model's top-$K$ predictions from the provided candidate set.

\paragraph{HitRate@K.}
Hit Rate measures whether the correct item appears among the top-$K$ predictions:
\[
\text{HitRate@K}(u)
= \mathbb{I}\{G_u \in \{\hat{y}_{u,1}, \ldots, \hat{y}_{u,K}\}\}.
\]
The final HitRate score is the average over all users.

\paragraph{NDCG@K}
Since each user has exactly one relevant (clicked) item, NDCG@$K$ simplifies to:
\[
\text{NDCG@K}(u)
= \sum_{k=1}^{K}
\frac{\mathbb{I}\{\hat{y}_{u,k} = G_u\}}{\log_2(k+1)}.
\]
We average NDCG@10 over all users.

\paragraph{Preliminary-round leaderboard score and choice of $K$.}
We report both HitRate@10 and NDCG@10. The official leaderboard ranks teams according to a weighted combination:

\[
\text{Score}_{\text{prelim}}
= 0.31 \cdot \text{HitRate@10}
+ 0.69 \cdot \text{NDCG@10}.
\]

The coefficients were calibrated on an internal pool of baseline models so that HitRate@10 and NDCG@10 contribute roughly equally to the final score:  
we first trained and evaluated multiple models, computed the average values of the two metrics and then chose the weights so that the two terms had comparable magnitude at these averages. 
We also empirically compared $K=10$ and $K=100$ on the same set of models and found that $K=10$ yields a larger coefficient of variation across systems, i.e., more diverse scores and clearer separation between teams. 
For this reason all official metrics are reported at $K=10$.

\subsection{Second-Round Evaluation Metrics}
\label{sec:eval-second}

\subsubsection{Weighted Hit Rate and NDCG}

In the second-round evaluation, we extend the relevance definition to incorporate both clicks and conversions. 
Weighted metrics, \ie, $w$-HitRate@10 and $w$-NDCG@10 are adopted by assigning different weights to different behavior types.

Let $G_u$ denote the set of ground-truth target items for user $u$ (including both clicks and conversions), and let $\hat{y}_{u,1}, \ldots, \hat{y}_{u,K}$ be the ranked list of items predicted by a model from the candidate set. 
We define a relevance weight function $w(i)$ that depends on the action type associated with item $i$:

\[
w(i) =
\begin{cases}
0, & \text{if } i \text{ is an exposure only}, \\
1, & \text{if } i \text{ is a click}, \\
\alpha, & \text{if } i \text{ is a conversion},
\end{cases}
\]

where $\alpha = 2.5$ to reflect the higher value of conversions.

The weighted HitRate@$K$ for user $u$ is:

\[
w\text{-HitRate@K}(u)
= \frac{\sum_{k=1}^{K} w(\hat{y}_{u,k}) \,
\mathbb{I}\{\hat{y}_{u,k} \in G_u\}}
{\sum_{i \in G_u} w(i)},
\]
where we average this quantity over all users.

For the weighted NDCG@$K$, we first compute the weighted DCG@$K$, then the weighted IDCG@$K$, and finally obtain the weighted NDCG@$K$ as their ratio.

\[
w\text{-DCG@K}(u)
= \sum_{k=1}^{K}
\frac{w(\hat{y}_{u,k}) \,
\mathbb{I}\{\hat{y}_{u,k} \in G_u\}}
{\log_2(k+1)}.
\]
We define the ideal $w$-DCG@$K$ for user $u$ by sorting items in $G_u$ in decreasing order of $w(i)$ and summing their discounted gains:

\[
w\text{-IDCG@K}(u)
= \sum_{k=1}^{\min(K, |G_u|)}
\frac{w(i_k^\star)}{\log_2(k+1)},
\]
where $i_1^\star, i_2^\star, \ldots$ is the ideal ordering. 
The weighted NDCG@$K$ is then:

\[
w\text{-NDCG@K}(u)
= \frac{\text{DCG@K}(u)}{\text{IDCG@K}(u)},
\]
where we again average $w-$NDCG@10 over users.

\subsubsection{Second-round Leaderboard Score}


We reuse the same weighting scheme as in the preliminary round, calibrated on internal baselines so that the contributions of $w$-HitRate and $w$-NDCG are approximately balanced (Section~\ref{sec:eval}).


In the second round, models must implicitly infer the underlying behavior type (click vs.\ conversion) when ranking candidate items. 
A mapping from user IDs to their corresponding ground-truth behavior types is provided in a separate file. 
The mapping is used solely within the evaluation pipeline to compute relevance weights and is not directly visible to submitted models.

%% file: 7.outcome.tex
\section{Challenge Summary}
\label{sec:outcomes}

\challenge{} attracted over 8{,}440 registered participants from nearly 30 countries and regions. In total, more than 140 universities outside mainland China (including institutions in Hong Kong, Macao, and Taiwan) and over 340 universities in mainland China took part in the competition, leading to about 4{,}600 active participants organised into more than 2{,}800 teams. 
Among all participating institutions, the five universities with the highest number of registrations were the University of Science and Technology of China (USTC), Tsinghua University, the University of the Chinese Academy of Sciences (UCAS), Zhejiang University, and Fudan University.

Here we briefly summarise the key modelling ideas of the top three teams and the winner of the Technical Innovation Award.

\paragraph{First place.}
The winning team built a multi-modal auto-regressive generative recommendation model on top of a dense Qwen backbone. 
They introduced a per-position action-conditioning mechanism~\citep{badrinath2025pinrec} that modulates token representations according to the action type using a combination of gated fusion, FiLM layers~\citep{perez2018film} and attention biasing, allowing the model to disentangle the semantics of different behaviors. 
They further engineered a hierarchy of time features capturing absolute timestamps, relative gaps and session structures (request, session, cross-day visit session), and encoded periodicity with multi-frequency Fourier features. 
To better represent long-tail items, they applied residual quantized $k$-means (RQ-KMeans)~\citep{luo2025qarm} to multi-modal embeddings to generate semantic IDs, combined with a random-$k$~\citep{fu2025forge} strategy to regularize training. 
A hybrid Muon\citep{jordan6muon} + AdamW optimizer~\citep{loshchilov2019decoupled} with a static-shape, GPU-friendly contrastive InfoNCE loss~\citep{oord2018representation} and large negative banks was applied during training. 
At inference time, they performed end-to-end generation of user vectors followed by ANN retrieval, achieving a favorable trade-off between performance and resource usage.

\paragraph{Second place.}
The runner-up solution adopted an encoder–decoder architecture. The encoder network used multiple gated MLPs~\cite{yang2025qwen3} to learn representations of users, items, and interaction sequences. The encoded context was further enriched by graph neural networks~\cite{xu2020inductive} operating on sampled neighborhoods from the user–item interaction graph.The decoder network was an improved SASRec-style~\cite{kang2018sasrec} Transformer that generated a ``next embedding'' representing the user’s future interest. The Transformer was configured with a 2048-dimensional hidden size, 8 layers, and 8 attention heads per layer. To capture semantic information from multi-modal context, the model employed an SVD-based residual-quantized -means (RQ-KMeans)~\cite{luo2025qarm} scheme to construct discrete semantic IDs. Following the practice of PinRec~\cite{badrinath2025pinrec}, it also performed conditional generation by encoding the behavior type of the next item during generation. Training followed a two-stage procedure: pre-training on exposure interaction events and fine-tuning on click and conversion events, with InfoNCE used as the learning objective. At inference time, the output embeddings of the decoder were used for ANN retrieval, followed by a post-processing step that filtered out items the user had previously interacted with.

\paragraph{Third place.}
The third-place team adopted a decoder-only Transformer for generative recommendation. For input features, they incorporated sparse user and item attributes together with rich temporal signals (e.g., absolute timestamps, relative time gaps). Additionally, following the design in PinRec~\citep{badrinath2025pinrec}, they introduced the next action type as an explicit conditioning signal, enabling the model to predict the next item under a specified behavior (e.g., exposure, click, conversion).

The model was trained with an InfoNCE loss~\citep{oord2018representation}, and training efficiency was further improved via AMP mixed-precision and static graph compilation. A core contribution of their approach was a systematic study of scaling laws~\citep{kaplan2020scaling} for generative recommendation, which yielded practical guidance for scaling large generative recommendation models under compute and memory budgets. Specifically, from a model perspective, they explored scaling along three aspects: (i) the number of negative samples used in the contrastive loss, (ii) model capacity (Transformer depth and hidden width), and (iii) the dimensionality of item-ID embeddings in the input representation.

Notably, they scaled the per-batch number of negatives up to 380K and observed substantial performance gains. Overall, these results reinforced the perspective that, for generative recommendation, performance was often driven more by scale than by intricate model design.

\paragraph{Winner of Technical Innovation Award.}
The awarded team proposed a decoder-only generative model that jointly modeled \emph{the user's next interested item} and \emph{the user action on the item}, with a unified training objective combining both semantic-ID generation loss and action prediction loss. 
This design explored an integrated paradigm which unified generative retrieval and ranking within a single model. 
It incorporated several state-of-the-art components, including FlashAttention~\citep{dao2022flashattention}, SwiGLU feed-forward networks, RMSNorm, RoPE ~\citep{su2024roformer}, and a DeepSeek-V3-style Mixture-of-Experts~\citep{liu2024deepseek}. 
The semantic-ID construction module introduced two key innovations: 
(i) a dedicated decoder-only transformer with InfoNCE loss for extracting collaborative embeddings of items, and 
(ii) a collision-resolution mechanism for the second-level semantic codes, which automatically searched for the next-closest cluster center when a code collision occured. 
On the feature-engineering side, the model leveraged not only the original sparse user/item features and multi-modal item representations, but also item popularity statistics across multiple temporal windows, as well as other discrete and continuous time features. 
For training and inference optimization, they adopted mixed-precision training, separate sparse/dense optimizer updates, grouped GEMM, and KV cache acceleration, resulting in substantial improvements in both efficiency and scalability.

%% file: 8.conclusion.tex
\section{Conclusion}

We have presented the TencentGR datasets and the \challenge{} centred on all-modality generative recommendation for advertising.
By releasing large-scale de-identified user sequences with rich multi-modal features, carefully designed evaluation metrics and a strong baseline, we hope to provide a valuable benchmark for the emerging field of generative recommendation.
We believe that the combination of realistic industrial data, a clearly defined generative task and a diverse set of strong baselines and solutions will catalyse further research on all-modality generative recommendation.

\section{Acknowledgement}

We want to express our sincere gratitude to the following individuals (in alphabetical order by surname)  for their invaluable contributions: 
Tao Guo, Meng Jin, Yue Liu, Lei Mao, Yuan Wang, Wenxiu Xue, Yuan Xie, Zhengwei Yang, Huanyu Yuan, Yuxin Zheng.

We gratefully thank the following individuals (in alphabetical order by surname) for their contributions on the challenge testing:
Shuchen Cai, Boqi Dai, Weitao Deng, Kai Fu, Hengxin Gao, Xuxuan He, Bin Hu, Yuyang Huang, Junbang Huo, Hanyong Li, Siao Li, Yuxuan Lin, Hongli Liu, Jiahua Luo, Chengyuan Mai, Chunwen Pan, Peiwen Pan, Wanqing Peng, Sikai Ruan, Chaoqun Su, Zixuan Su, Wangbin Sun, Hongbo Tang, Jinyuan Wang, Yuxin Wang, Zixiao Wang, Yupeng Wei, Jianbing Wu, Na Xu, Wei Xu, Zeen Xu, Fengyu Yang, Haixin Yang, Muzhen Yang, Shichen Yang, Yi Yang, Xiangxin Zhan, Jiangtao Zhang, Rimin Zhang, Xinyue Zhang, Yongqi Zhou, Jie Zhu.